# Exchange Coupling and Decoupling in a Nanocomposite Spin System[1]

Y. Z. Shao[2]   W. R. Zhong   T. Lan   R. H. Lee

(Department of Physics, Sun Yat-sen University, Guangzhou 510275, P. R. China)

**Abstract:**   We studied the exchange coupling and decoupling occurring in a nanocomposite spin system based on a 3D Heisenberg model by means of Monte Carlo numerical computation simulation. Different from conventional micromagnetism approach which usually adopts finite elements method to compute, in a top-down way, the magnetic property of micromagnetic ensemble in micron even nanometer scale, our approach in this paper is peculiar to the structure of a complex spin lattice, i.e. two species of spins building up from single spin to cluster spins in a bottom-up way. The simulation revealed the influence of exchange coupling constant $J_{ab}$, the size of cluster spins $d$ and system reduced temperature $t$ upon the exchange coupling and decoupling between component spin phases of a nanocomposite magnets, respectively. Smaller value of $J_{ab}$, larger $d$ and lower temperature $t$ usually lead to the decoupling of originally exchange-coupled component phases and the occurrence of a characteristic two-stage shoulder with an inflexion on the demagnetization curve. The results reported in this paper are of, to some extent, universality and applicable to other dual-phase magnetic systems since our simulation simply focus on a pure duplex spin system rather than a specific material and all physical variables were treated in a reduced form.

## 1. Introduction

The intriguing characteristics of nanocomposite magnets have attracted the intensive attentions of physicists and magnetic material researchers in last decade, and aroused numerous studies on both fundamental micromagntism and the practical application of nanomagnetic materials. One can trace some typical applications of nanocomposite magnets in soft and hard magnetic materials [1~6]. The exchange coupling and decoupling of cooperative magnetic moments between component magnetic phases is key factor underlying the distinctive behavior of nanocomposite magnets. According to the characteristics of component magnetic phases, nanocomposite magnets could be classified into two typical cases: first, identical in magnetic moment but different in crystalline structure, such as crystalline α-Fe plus matrix amorphous Fe in dual-phase Finemet [2,7]; second, total heterogeneous magnetic components, such as Nd(Pr)FeB plus nanoscaled α-Fe[3,5,8,9]. No matter which combination of magnetic phases above is concerned, the exchange coupling between magnetic phases plays an important role in improving the magnetic behavior of nanocomposite magnets. For instance, the exchange coupling gives rise to the raising Curie temperature of remnant amorphous matrix of excellent soft magnet Finemet [2] as well as the increasing energy product of hard nanocomposite magnet Nd(Pr)FeB plus α-Fe[3,5,8,9]. Generally, the micromagnetic approach based on phenomenological mean-field approximation is theoretical basis to tackle those micromagnetic systems, and finite element method (FEM) is usually employed in a top-down way to compute the magnetic property of micromagnetic systems [5,8,9]. Some problems, however, probably arise in two aspects below. Firstly, it is well-known that mean-field theory undoubtedly results in considerable error nearby Curie temperature owing to the weakness of the phenomenological theory itself. Secondly, the traditional top-down micromagnetic computation by means of FEM probably causes somewhat errors and becomes implausible when the grain size of materials is within the range of nanoscale. Skomski and Aharoni have ever respectively expounded some typical misinterpretations and mistakes in theoretical approach of some micromagnetics and one may refer to their review articles [1,10]. Moreover, most of micromagnetic computations by FEM were related to a specific material [1,5,8,9] and material parameters, even in bulk state of material, were used. Actually, the universality and applicability out of these micromagnetic computations are limited. In addition, the different microstructural patterns such as magnetic domain wall and boundary, if adopted, also complicates the exchange coupling and decoupling of two magnetic phases, and the results are usually microstructure dependent even though some of magnetic behaviors are microstructure independent.

---

[1] Supported by the Natural Science Foundation of Guangdong Province, P.R.China (No.031554)

[2] Correspondent author, email: stssyz@zsu.edu.cn

In this paper, we simulated the exchange coupling and decoupling of a nanocomposite spin system on the basis of 3 dimensional Heisenberg model by means of Monte Carlo numerical computation. Unlike those conventional FEM micromagnetism approaches, our approach is peculiar to the frame of a complex spin lattice, i.e. two species of spins building up from single spin to cluster spins in a bottom-up way. Without any bulk material parameters included, our simulation was carried out simply using single- or cluster-spin value and all physical variables were treated in a reduced form. For the sake of simplicity, direct contact of two sorts of magnetic phases on a single layer of coherent interface was considered in order to get rid of such side effects as interfacial strain energy and domain wall energy caused by different breadth of domain wall. In this paper, we concentrated on the impact of three factors upon the exchange coupling and decoupling. They are the constant of exchange coupling between two magnetic phases $J_{ab}$, cluster-spin size $d$ and the reduced temperature of system $t$, respectively. Among the three factors, the influence of component α-Fe grain size on the magnetic property of hard nanocomposite magnet Nd(Pr)FeB plus α-Fe was studied by FEM micromagnetic calculation [5] and experimental investigations in detail [3,8,9]. As for the influence of grain size on the magnetism in a soft nanocomposite magnets, one may refer to these references [1,2,4] for further information.

## 2. Descriptions of Model and Computational Algorithm

Enlightened by Fisch's work in simulation of the behavior of a duplex-magnetic phase spin system [11], we modified the Hamiltonian of a classical Heisenberg spin system (the first term in Eqn.1.1) by introducing both determinant uniaxial single-ion anisotropy parallel to the z-axis (the second term) and random uniaxial anisotropy (the third term) in the form as shown in Eqn.1.1. The fourth term is Zeeman energy with an external driving field parallel to either Z-axis direction <001> or diagonal <111>.

$$\hat{H} = -(J_{NN} + J_{NNN}) \sum_{<i,j>} (\vec{S}_i \cdot \vec{S}_j) - A \sum_{i'} (\vec{S}_{i'}^z)^2 - D \sum_{i''} ((\vec{S}_{i''} \cdot n_{i''})^2 - 1) + \mu H \sum_i \vec{S}_i^Z \qquad (1.1)$$

where $\quad \vec{S}_i \cdot \vec{S}_j = c_x S_i^x S_j^x + c_y S_i^y S_j^y + c_z S_i^z S_j^z$ \hfill (1.2)

Symbol $S_i$ and $S_j$ in Eqn.1.1 represent the spin at site $i$ and neighboring site $j$,

respectively, within the lattice. $S^x$, $S^y$ and $S^z$ in Eqn.1.2 denote, in turn, the projections of spin $S$ along x-, y- and z-axes. The spin-exchange constants $J_{NN}$ and $J_{NNN}$ signify the interaction of site $i$ with its nearest neighbors (NN) and the next-nearest neighbors (NNN), respectively. Depending on whether $J_{NN}$ or $J_{NNN}$ is selected, the summation $\Sigma_{<i,j>}$ symbolizes the sum over NN or NNN site pairs, respectively. Parameters $c_x$, $c_y$ and $c_z$, ranging from 0 to 1, are anisotropy constants of spin exchange. The isotropic Heisenberg spin system corresponds to the case when $c_x = c_y = c_z =1$ and this is the case we handle in current project. On the other hand, if $c_x= c_y= 0$ and $c_z =1$ the Hamiltonian describes the anisotropic Ising spin system. Parameter $\mu$ is a constant related to magnetic moment. Parameter $A$ and $D$ are the anisotropy constants of two distinctive spin sites. Four types of energy terms in the components of the Hamiltonian in Eqn.1.1 account for the spin exchange energy of coupling between single-spins or cluster-spins, determinant uniaxial single-ion anisotropy energy, random uniaxial anisotropy energy and interaction energy between spin and external field, respectively. The first term includes all possible spin sites throughout the entire 3D lattice and the first term governs the spontaneous magnetism of system. The spins of the second and the third terms, however, only occupy those sites belonging to either cluster spin sites (site $i'$) of crystallite part with a determinant orientation to a easy axis (Z axis) or single-atom spin sites (site $i''$) of matrix part with numerous random orientations, respectively. The ***$n_{i''}$*** denotes a unit vector independently chosen for each site $i''$ with a random local easy direction which varies from site to site. Apparently, site $i'$ and site $i''$ are of magnetic anisotropy and isotropy in nature and called the anisotropic site (*phase a*) and isotropic site (*phase b*) respectively in this paper. Assuming the volume fraction of cluster-spin site $i'$ (*phase a*) and single-spin site $i''$ (*phase b*) to be *1-x* and *x* respectively, we could construct a complex lattice site with a continuous variation of *x* value. Depending upon the condition of simulation, it is feasible for us to adjust the size *d* of cluster-spin site $i'$ and its volume fraction *1-x* in computation. The simulation software will check automatically the compatibility of selected values of *x* and *d* in order to avoid any cavities emerging within 3D lattice due to the mismatch of *d* and *x*, e.g. large *d* and *x*. Concerning the configuration of specific spin alignment, three kinds of spin exchange constants were defined for this complex lattice respectively, i.e. $J_{aa}$, $J_{bb}$ and $J_{ab}$, signifying the direct coupling intensities and cooperative capability among those spins within *phase a* and *phase b* themselves ($J_{aa}$, $J_{bb}$) as well as the exchange coupling intensity of interface spins between *phase a* and *phase b* ($J_{ab}$). According to the local spin-site and spin–pairs in 3D lattice scanned by Monte Carlo simulation, the software program can identify the local microenvironment of lattice and decide which one of three exchange constants to be selected. Considering the situation of a typical duplex hard-soft magnetic phase system, we set the anisotropy constant *A* and *D* in Eqn.1.1 to a fixed ratio A/D=1000 in most of cases, namely a strong magnetic anisotropic crystallite *phase a* against a nearly isotropic *phase b*. Through the combinations of the values of *S, μ, A* and *D* of *phase a* and *phase b* in Eqn.1, we could also simulate the magnetic behaviors of a variety of duplex magnetic phase systems, such as duplex hard-hard magnetic system or soft-soft magnetic one. The spontaneous magnetism of a duplex nanomagnetic phase system could also be worked out by our simulation when driving field *H* = 0, and we reported it in our recent paper [12]. In light of the theory of superparamagnetism that all single-atom spins within a cluster align in identical direction, we could figure out the spin of a cluster-spin as $S_{Cluster} \sim d^3 S_{atom}$, where $S_{atom}$ is the spin of single-atom. The 3D lattice (size *N* with $N^3$ sites) of simulation is comprised of numerous small basic cubic sub-lattices and cluster-spin size *d* is measured in the unit of a basic cubic cell consisting of eight corner atoms.

The most fundamental reduced parameter $k_BT_C/J_{NN}$, the ratio of the critical temperature against the exchange interaction, was calculated using Eqn.2 [13],

$$k_BT_C/J_{NN} = 5(R - 1)[11S(S + 1) - 1]/96 \qquad (2)$$

where $R$, $S$, $k_B$ and $T_c$ are the number of the nearest neighbors, lattice spin, Boltzmann constant and critical temperature, respectively. The interaction of site $i$ with the next-nearest neighbor $J_{NNN}$, which drops exponentially with respect to the distance between lattice sites[14], can be determined from $J_{NN}$ and is usually taken as 0.1~0.25 $J_{NN}$ [15]. In current paper, we take $J_{NNN}$ to be 0.2 $J_{NN}$ and critical temperature $T_c$ was set to an ambient 293K.

The standard Metropolis criterion [16] was employed in our Monte Carlo simulation. The simulation proceeded by sweeping every lattice site in sequence for a number of repetitions (Monte Carlo steps, MCS) and the statistical averages of the magnetic properties concerned were computed over ten independent simulations. Standard tests were performed to verify whether equilibrium under the prescribed condition was attained. All simulations were performed on a three-dimension lattice with a periodic boundary condition and lattice size $N$=60~100. In our simulation the sweeping times counted up to $10^5$ MCS.

With $\vec{S}_{i'}$ and $\vec{S}_{i''}$ denoting the magnetic spins occupying site $i'$ and site $i''$ respectively, the magnetization $m$ averaged over all lattice sites is given by

$$\vec{m} = \frac{1}{N^3} \sum_{i \in i'+i''} (\vec{S}_{i'} + \vec{S}_{i''}) \qquad (3)$$

The demagnetization curve and hysteresis loop of duplex nanomagnetic spin system could be obtained by simulating the variation of magnetization $m$ against driving field $H$.

## 3. Results

Occurrence of a two-stage shoulder with a distinctive inflexion on demagnetization curve shows the exchange decoupling of duplex magnetic phase system. The magnetic moments of two component magnetic phases, owing to thorough exchange coupling between them, rotate coherently in accordance with driving field and a smooth demagnetization curve is observed. Due to exchange decupling, the original duplex nanomagnetic system turns into a common macroscopic mixture and different coercive forces comes into being. Figure 1 shows the typical demagnetization curves and differential susceptibilities of duplex nanomagnetic systems with a fixed cluster spin size $d$ in small and large exchange coupling constant $J_{ab}$ values. The increase in $J_{ab}$ intensifies the exchange coupling straightforwardly and weakens the two-stage shoulder of demagnetization curve and the peak of differential susceptibility. Similar situation also exists in the variation of cluster spin size $d$ while $J_{ab}$ holds invariably, as displayed in figure 2. The perfect exchange coupling occurs easily in a dual namomagnetic system with little size component phases. Necessarily note that the critical values of $J_{ab}$ and $d$ at which exchange decoupling takes place are variables dependent upon other parameters of component phases.

The temperature dependence of exchange decoupling of a duplex nanomagnetic system is more complicated than the situations of size and exchange coupling constant. We investigated in detail the temperature dependence of exchange decoupling of system in various cases of component phases. Plotted in figure 3 were the magnetic hysteresis loops when $\mu_a=\mu_b$ and $X$=0.2 and 0.8, respectively. Among the figure 3, figure 3a and 3b correspond to the situation that phase $a$ and phase $b$ possess an identical spin value, i.e. $S_a = S_b$. The hysteresis loops of $S_a \neq S_b$ were exhibited in figure 3c and 3d. The counterparts of $\mu_a \neq \mu_b$ were presented in figure 4. The results of figure 3 and figure 4 demonstrate the general feature of the exchange decoupling of a duplex nanomagnetic system as the temperature of system declines. The hysteresis loop of duplex nanomagnetic system dominated by anisotropic *phase a* ($X$=0.2) yields a larger coercivity and takes on a more square shape in comparison with that by isotropic *phase b* ($X$=0.8). Considering the situation when *phase a* is completely identical with *phase b* in their magnetic moments, namely $\mu_a=\mu_b$ and $S_a=S_b$, the magnetic dissimilarity contributed by two phases is originated simply from their differences on the parameter $A$ and $D$ in Hamiltonian eqn 1.1. And in this situation, the exchange decoupling never occurred even at reduced temperature as low as $t$ = 0.05 if isotropic *phase b* dominates, $X$=0.8 as shown in figure 3b. The distinctive two-stage shoulder of exchange decoupling, however, appeared even at temperature as high as $t$ = 0.3 ~ 0.5 and $X$=0.8 while $S_a \neq S_b$ as displayed in figure 3d. Interestingly, notice that the inflexions of two-stage shoulder of exchange decoupling keep invariably as temperature changes. When $\mu_a \neq \mu_b$, exchange decoupling happened merely at a certain temperature, e.g. at $t$ = 0.2 in figure 4b, contrast to preceding those that exchange decoupling never disappeared once it takes place below a specific temperature. Figure 4b indicates that exchange decoupling does not occur invariably at lower temperature.

## 4. Discussion

The exchange coupling has a positive effect on the characteristic of nanocomposite magnetic system and exchange coupling and decoupling exist in both demagnetization and spontaneous magnetization [1]. Theoretically, exchange coupling between two magnetic phases is a short-range interaction less than 10 nanometers for most of magnetic systems [1]. The extent of exchange coupling $l_{ex}$ could be expressed as below [1]:

$$l_{ex} = \sqrt{\frac{A_{ex}}{\mu_0 M_s^2}} \quad \quad \quad 4.1$$

$$A_{ex} = \frac{JS^2}{l} \quad \quad \quad 4.2$$

where $A_{ex}$, $Ms$ and $l$ are exchange coupling stiffness, spontaneous magnetization and dimension of lattice cell, respectively. Apparently, the extent of exchange coupling $l_{ex}$ decline if the spontaneous magnetization $Ms$ increases. Because of declining $l_{ex}$, some of the originally exchange-coupled grains of component phases degrade into partially even completely decoupled status, and nanocomposite magnets turns into a trivial macroscopic mixture. The magnetic behavior of system is a simple addition of that of each component magnetic phases themselves.

It is easy to understand the reason why small $J_{ab}$ and large $d$ bring about the exchange decoupling between component magnetic phases. Previous works have gained an insight into the influence of grain size on nanocomposite hard magnets Nd(Pr)FeB plus α-Fe [3,5,8,9]. What we emphasize here is the critical size $d_c$ at which exchange decoupling takes place is controllable through adjusting both basic parameters relative to element species such as $\mu$, $S$, $J_{ab}$ and other parameters relative to microstructure such as $A,D,X$ as well as system temperature $t$。 It is possible to avoid the occurrence of exchange decoupling by systematically adjusting relative parameters above.

The study on exchange decoupling owing to temperature is insufficient compared with the investigations on gain size. The explanation for the temperature dependence of exchange decoupling, based on past FEM micromagnetic calculation, is not satisfactory in that the variation of such microstructural factor as the width of Bloch domain wall $\delta_B$ with temperature was solely regarded as the root of exchange decoupling [8]. In the light of FEM micromagnetism, exchange coupling takes effect only when exchange-couple length $l_{ex}$ is equivalent to Bloch wall width $\delta_B$ in scale. As $\delta_B$ gets smaller with decreasing temperature due to $\delta_B \sim 1/\sqrt{K}$ and anisotropy constant $K$ increasing with decreasing temperature, the grain size necessary for a complete exchange coupling is shifted to smaller values so that some of the bigger grains behave partly or even completely decoupled from the neighboring grains and reverse independently [8]. In our current new approach there is not any domain wall width concerned and our simulation manifests the appearance of exchange decoupling at lower temperature as well. We think the interpretation based upon the variation of exchange-couple length $l_{ex}$ due to spontaneous magnetization $Ms$, as revealed in Eqn.4, more essential and reasonable than that upon domain wall width $\delta_B$ in their effects on low-temperature exchange decoupling. The early explanation of wall width relative to exchange coupling length seems somewhat implausible in the situation of soft magnets as pointed out in reference [1]. For an excellent soft magnet, negligible anisotropy constant brings in a very great wall width, i.e. $K \to 0$ and $\delta_B \to \infty$, which would realize exchange coupling on a macroscopic scale. It is, however, impossible to achieve exchange coupling on a macroscopic scale because exchange coupling between component phases itself is characteristic of short-range and localized around their interface.

We can not give a full interpretation why exchange decoupling occurs only at an intermediate temperature when $\mu_a \neq \mu_b$, as shown in figure 4b. It probably, we speculate, has something to with the different dynamical response of two component phases to driving field. Hysteresis loop is characteristic of nonlinear and dynamical nonequilibrium, and influenced by the nature of component phases, the frequency and amplitude of driving field and the temperature of system simultaneously [16~21]. Parameter $\mu$ is a constant associated with magnetic moment and it affects Zeeman energy in Hamiltonian Eqn1.1. The dissimilar sensitivity of Zeeman energy to external field, due to diverse $\mu$ values of component phases, is probably most prominent under a certain condition e.g. at a certain temperature, giving rise to different dynamical responses to external field and leading to a two-stage shoulder on demagnetization curve. However, the emergence of two-stage shoulder under a specific condition may be disappeared at another temperature or external field since what we obtained in current simulation is dynamical other than static hysteresis loop.

The phenomenon of exchange coupling and decoupling is simply small part of features of duplex nanomagnetic system. It is necessary to execute more systematical investigations on the feature of dynamical hysteresis loop under various conditions.

**5. Conclusions**

The phenomenon of exchange coupling and decoupling between component phases is of special consequence to the theory of nanomagntism and the practical application of nanocomposite magnets also. Following are some important summaries drawn from our current simulation for this paper. First in methodology, we presented in our project a new approach to a duplex nanomagnetic system. Our new approach is based on Heisenberg exchange model and was successfully carried out using Monte Carlo simulation on a complex lattice of single- and cluster-spin developing in the bottom-up way, in contrast with conventional micromagnetic approach based on mean-field theory and calculating in top-down finite element method. Secondly, three kinds of important factors that impact on exchange decoupling, exchange coupling constant $J_{ab}$, grain size $d$ and temperature $t$, were studied in detail and the results are satisfactory. Exchange decoupling is easily triggered by little $J_{ab}$, large $d$ and low temperature $t$. Our simulation also proved that the decrease in exchange coupling length $l_{ex}$ causes exchange decoupling rather than the decrease of Bloch domain wall width $\delta_B$ as previous reference suggested. There certainly exist exchange coupling and decoupling in a duplex nanomagnetic system without domain wall width. The occurrence of exchange coupling and decoupling is also controllable by adjusting the composition and microstructure of component phases, temperature of system and external field. Thirdly, some new features probably appear, such as that exchange decoupling occurs only at an intermediate temperature observed in current simulation, owing to the nature of nonlinear and nonequilibrium dynamical response of different component phases to driving field.

Acknowledgements
This project is financially supported by the Natural Science Foundation of Guangdong Province, P.R.China at grant No.031554. Authors are gratitude to Professor G. M. Lin for his valuable suggestion and discussion.

Reference
1   R. Skomski, *J.Phys:Condens Matter*, 15(2003)R841
2   A. Hernando, I. Navarro and P. Gorria, *Phys.Rev. B.*, 51(1995)3281.
3   G. C. Hadjipanayis, *J. Mag. Mag. Mater*. 200(1999)373.
4   U. Bovensiepen, F. Wilhelm, P. Srivastava, P. Poulopoulos et al., *Phys. Rev. Lett.*, 81(1998)2368.
5   T. Schrefl, R. Fischer, J. Fidler and H. Kronmuller, *J.Appl.Phys.*, 76(1994)7053; *Phys.Rev.B*,49(1994)6100
6   R. Skomski and D. J. Sellmyer, *J.Appl.Phys.*, 89(2001)7263; 87(2000)4756
7   Y. Yoshizawa, S. Oguma and K .Yamauchi, *J.Appl.phys.*, 64(1988) 6044.
8   D. Goll, M. Seeger and H. Kronmuller, *J. Mag. Mag. Mater.*, 185(1998)49.
9   H. Kronmuller, R. Fischer, M. Seeger and A. Zern, *J. Phys. D:Appl. Phys.*, 29(1996)2274.
10  A. Aharoni, *Physica B*, 306(2001)1.
11  R. Fisch, *Phys. Rev. B.*,58(1998)5684.
12  Y. Z. Shao, W. R. Zhong, G. M. Lin and X. D. Hu, archXiv-cond
13  G. S. Rushbrooke and P. J. Wood, *Mol. Phys.*, 1(1958)257.
14  D. Wagner, *Introduction to the theory of magnetism*, Pergamon Press, New York, 1972. p153.
15  J. Mlodzki, F. R. Wuensch and R. R. Galazka, *J. Mag. Mag. Mater*, 86(1990)269.
16  K. Binder and D. W. Heermann, *Monte Carlo simulation in statistical physics*, Springer, Berlin, 1992. p14.
17  B. K. Chakrabarti and M. Acharyya, *Rev. Mod. Phys.*,**71**(1999)847.
18  M. Acharyya and B. K. Chakrabarti, *Phys. Rev. B.*, **52**(1995) 6550.
19  S. W. Sides, P. A. Rikvold, and M. A. Novotny, *Phys. Rev. E*, **59** (1999)2710; *Phys. Rev. E*, **57**(1998) 6512.
20  M. Rao, H. R. Krishnamurthy and R. Pandit *Phys Rev B*, **42**(1990) 856
21  T. Tome and M. J. Oliveira, *Phys. Rev A*, **41**(1990)4251
22  Y. Z. Shao, J. K. Lai, C. H. Shek, G. M. Lin and T. Lan *Phys. Stat. Sol. B* **232**(2002) 330

Figure captions:

Figure 1. The demagnetization and differential susceptibility curves of a nanocomposite magnetic system with different exchange coupling constants $J_{ab}$.

Figure 2. The demagnetization and differential susceptibility curves of a nanocomposite magnetic system with different sizes of cluster spin $d$.

Figure 3. The magnetic hysteresis loops of nanocomposite magnetic system with $\mu_a=\mu_b$ in a smaller ($X=0.2$) and larger ($X=0.8$) volume fraction of random anisotropy spin sites at various reduced temperatures $t$. Among these figures, component phases $a$ and $b$ own either an identical or different spin values as shown in figure (a), (b) $S_a=S_b=0.5$ and (c), (d) $S_a=0.5$ $S_b=1.5$ respectively.

Figure 4. The magnetic hysteresis loops of nanocomposite magnetic system with $\mu_a \neq \mu_b$ in a smaller ($X=0.2$) and larger ($X=0.8$) volume fraction of random anisotropy spin sites at various reduced temperatures $t$. The other conditions similar to those of figure 3(c) and (d).

Figure 1

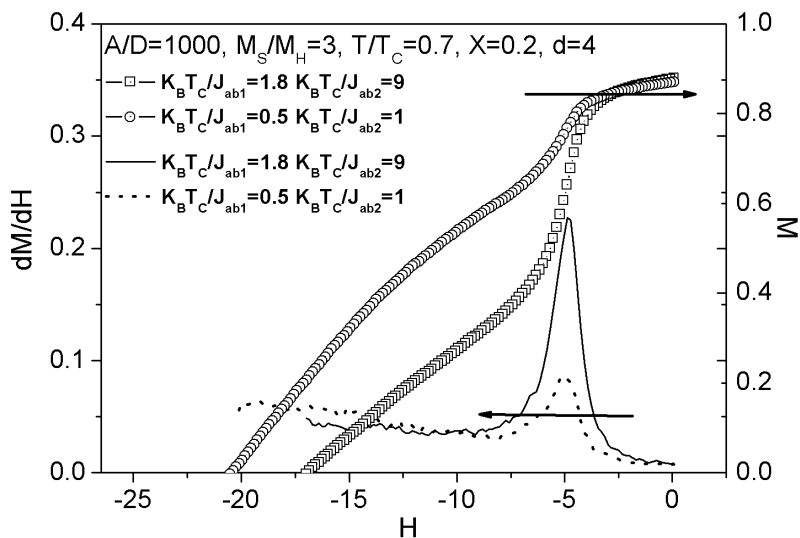

Figure 2

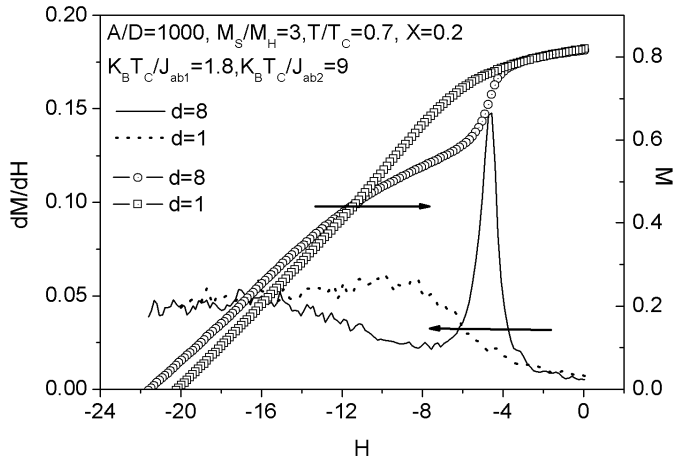

Figure 3

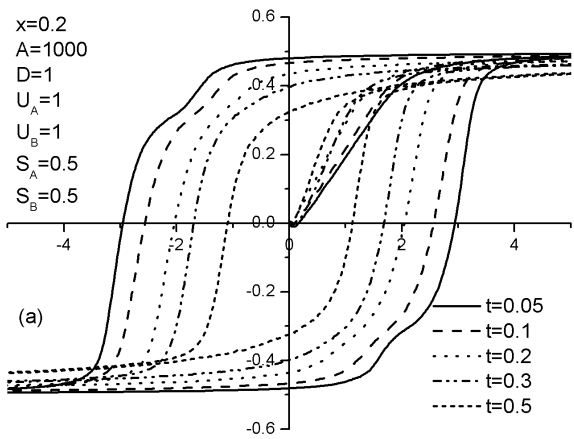

(a)

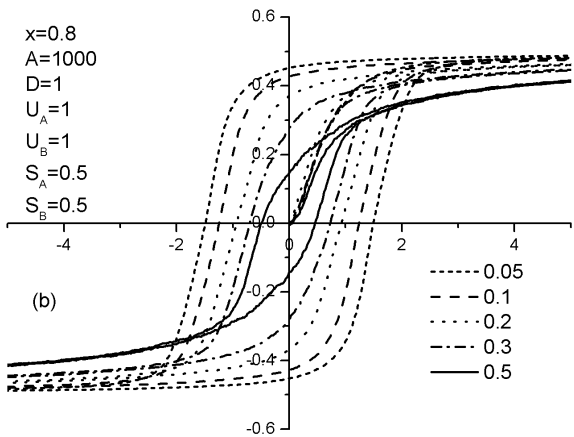

(b)

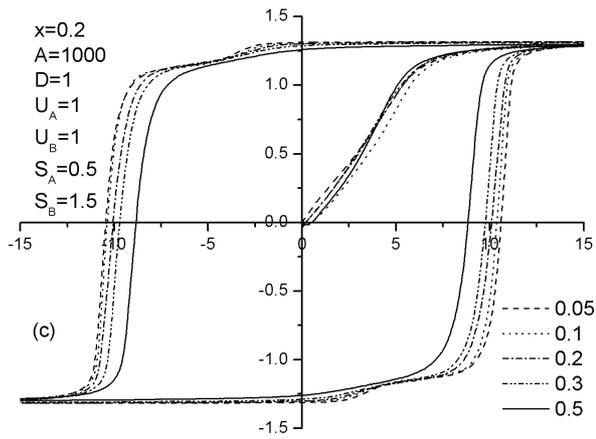

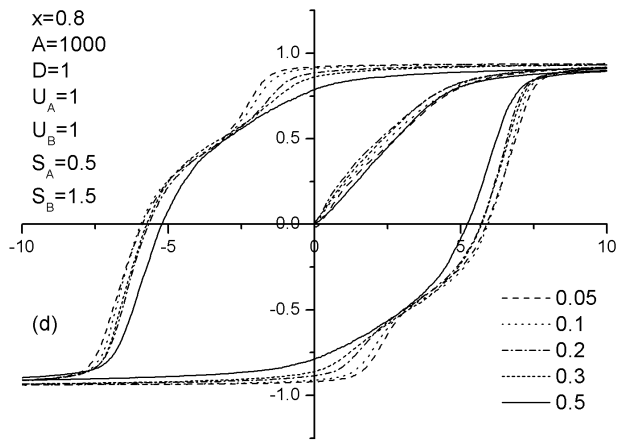

Figure 4

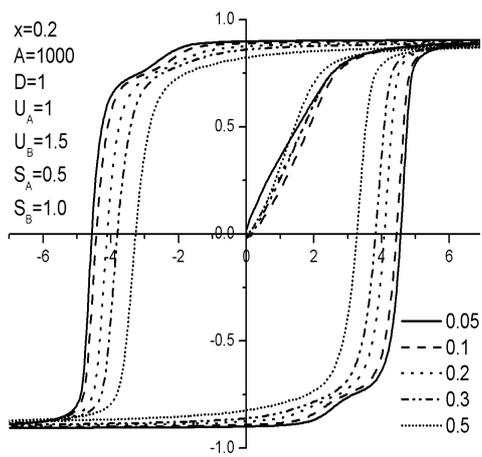
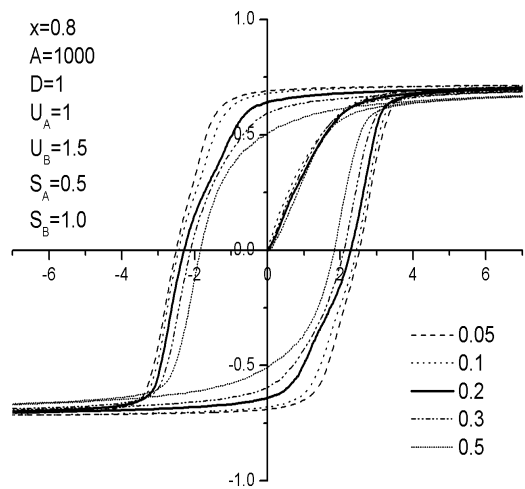